\documentclass[useAMS,usenatbib]{mn2e}
 \usepackage{amsmath}
  \usepackage{graphicx}
  \usepackage{subfigure}
  \usepackage{comment}
  \usepackage[usenames]{xcolor}
  \usepackage{bm}
  \usepackage{epstopdf}

\newcommand{\kepler}{{\em Kepler}}
\newcommand{\corot}{{\em CoRoT}}
\newcommand{\numax}{\mbox{$\nu_{\rm max}$}}
\newcommand{\Dnu}{\mbox{$\Delta \nu$}}

\newcommand{\muHz}{\mbox{$\mu$Hz}}

\newcommand{\half}{{\textstyle\frac{1}{2}}}
\def\note #1]{{\bf #1]}}

\title[Characterizing mixed modes in evolved stars]{Using the phase shift to asymptotically characterize the dipolar mixed modes in post-main-sequence stars }
\author[C.~Jiang, J.~Christensen-Dalsgaard and M. Cunha]{C.~Jiang$^{1,2}$\thanks{E-mail:jiangch53@mail.sysu.edu.cn}
 J.~Christensen-Dalsgaard$^{3}$ and M. Cunha$^{2}$.\\
 $^{1}$School of Physics and Astronomy, Sun Yat-sen University, 2 Daxue Road, Tangjia, Zhuhai, 519082, Guangdong Province,  China \\
$^{2}$Instituto de Astrofísica e Ci$\hat{e}$ncias do Espa\c{c}o, Universidade do Porto, CAUP, Rua das Estrelas, PT4150-762 Porto, Portugal \\
$^{3}$Stellar Astrophysics Centre, Department of Physics and Astronomy, Aarhus University, Ny Munkegade 120, DK-8000 Aarhus C, Denmark}

\pagerange{\pageref{firstpage}--\pageref{lastpage}} \pubyear{2016}

\def\LaTeX{L\kern-.36em\raise.3ex\hbox{a}\kern-.15em
    T\kern-.1667em\lower.7ex\hbox{E}\kern-.125emX}

\begin{document}

\label{firstpage}

\maketitle

\begin{abstract}
Mixed modes have been extensively observed in post-main-sequence stars by the \kepler\ and \corot\ space missions. The mixture of the p and g modes can be measured by the dimensionless coefficient $q$, the so-called coupling strength factor. In this paper we discuss the utility of the phase shifts $\theta$ from the eigenvalue condition for mixed modes as a tool to characterize dipolar mixed modes from the theoretical as well as the practical point of view. Unlike the coupling strength, whose variation in a given star is very small over the relevant frequency range, the phase shifts vary significantly for different modes. 
The analysis in terms of $\theta$ can also provide a better understanding of the pressure and gravity radial order for a given mixed mode. Observed frequencies of the \kepler\ red-giant star KIC 3744043 are used to test the method. The results are very promising.  
\end{abstract}


\section{Introduction}

Thanks to the very long duration and high precision of photometric space observation missions, such as \kepler\ \citep{bor08,bor10} and \corot\ \citep[e.g.,][]{bag06a, bag06b, mic08}, the study of mixed modes in post-main-sequence stars, specifically subgiants and red giants, became a highlight of asteroseismology \citep{bed10, bec11}. Not only because space missions allowed us the detection of these small amplitude modes, but also because these modes carry information about the stellar interior \citep{bec12} which is not directly observable. Mixed modes behave like pressure (p) modes in the outer region while they behave like gravity (g) modes in the core of stars that have evolved off the main sequence (MS) after the depletion of hydrogen in the central parts.

When a star has evolved past the MS and entered the sub-giant branch (SGB), the very condensed core raises the gravitational field and hence the buoyancy frequency, which opens up the possibility of the coupling between p and g modes. Although the theory of the coupling between p and g modes has been used in numerical computations for a long time, the research concerning observed mixed modes increased significantly only in recent years \citep{jcd12}. Through the analysis of \kepler\ data, \cite{mos12a} managed to estimate the mean coupling strength of modes in red-giant branch (RGB) stars as well as clump giants using the asymptotic relation discussed in \cite{unn89}. \cite{ben12} introduced a way to fit mixed modes in SGB stars and announced that the coupling strength of the dipolar mixed modes is predominantly a function of stellar mass and appears to be independent of metallicity. A further detailed analysis of asymptotic relations for dipolar mixed modes proves its usefulness for predications of some general parameters, like coupling strength and period spacings \citep{jiang14}. Properties of dipolar oscillations in luminous red giants have been studied by \cite{dzi12} who solved the equations for adiabatic oscillations based on asymptotic decomposition of the fourth-order system. In this paper, we continue the asymptotic analysis of mixed modes in post-main-sequence giants. The paper starts with the investigation of the eigenvalue condition for mixed modes, and then extends the discussion to some general properties of mode coupling and the use of the phase shift in the asymptotic descriptions.

\section{General properties of mixed modes in giant stars}
An oscillation mode can be specified by three quantum numbers: the radial order $n$, the degree $l$ and the azimuthal order $m$. In the absence of rotation, magnetic field, and other physical agents that may break the spherical symmetry of the star,  the oscillation frequencies are degenerate in $m$. This is the case that will be considered in this paper. At low degree the cyclic frequencies of high-order p modes are approximately given by \citep{tas80,gou93}
\begin{equation}
  \nu_{nl} = \frac{\omega}{2\pi}\simeq \Dnu_{\rm as} (n + \half l + \epsilon_{\rm p}) - d_{nl}.
        \label{eq:asyp}
\end{equation}
Here $\omega$ is the angular mode frequency and $\epsilon_{\rm p}$ is a frequency-dependent offset due to the large gradient of the cut-off frequency near the stellar surface. An estimate of $\epsilon_{\rm p}$ can be deduced from the effective temperature and the line widths of modes \citep{whi12} by ignoring its frequency dependence. In the asymptotic formulation, e.g. equation (5.8.31) in \cite{gou93}, there is a frequency-dependent second-order term that consists of two components. The first of these, denoted by $d_{nl}$ in equation~\eqref{eq:asyp}, is related to the small frequency separation that is sensitive to the conditions in the core of the star. The second component depends predominantly on the surface layers. Since $\epsilon_{\rm p}$ is also largely related to the surface layers, here we have opted to absorb the higher-order term into $\epsilon_{\rm p}$. Also
\begin{equation}
  \Dnu_{\rm as}=\left(2\int_{0}^{R}\frac{{\rm d} r}{c(r)}\right)^{-1}
      \label{eq:dnu}
\end{equation}
is the large frequency separation, which is the inverse sound travel time across the whole star. The integral is of the sound speed $c(r)$ between the centre of the star, at $r=0$, and the surface where $r = R$. 
%An observationally motivated asymptotic expansion of the pure pressure mode is given by \cite{mos13} including also a second-order term to account for the curvature of the observed radial oscillation pattern in the \'{e}chelle diagram.}. 
On the other hand, the oscillation period of a g mode also satisfies a simple asymptotic relation \citep{tas80}:
\begin{equation}
  \Pi_{nl}=\frac{2\pi}{\omega_{nl}} \simeq \Delta\Pi_l\left(n + \frac{1}{2} + \epsilon_{\rm g} \right),
  \label{eq:asyg}
\end{equation}
where $\epsilon_{\rm g}$ is again an offset and the period spacing is
\begin{equation}
  \Delta\Pi_l = \frac{2\pi^2}{L}\left(\int_{r_1}^{r_2}N\frac{{\rm d}r}{r}\right)^{-1}.
  \label{eq:dp}
\end{equation}
Here $L=\sqrt{l(l+1)}$, $N$ is the buoyancy frequency (the so-called \textit{Brunt-V\"ais\"al\"a} frequency) and the integral is over the cavity where the g-mode oscillation is trapped. For dipolar modes, the propagation regions of p and g modes are theoretically determined by the frequencies and are well discussed in \cite{bookas}. In principle, acoustic waves travel in the area where $\omega > S_l$, $S_l = L c/r$ being the \textit{Lamb} frequency, in the outer part of the star, while gravity waves are trapped in the region below the convective envelope where $\omega < N$ and typically such that $\omega \ll S_l$. In the case of an RGB star, the two regions are well separated by a so-called evanescent region, within which the modes have an exponential variation with depth.
The width of the evanescent region is related to the coupling strength of mixed modes that have oscillatory behaviour in both the acoustic- and the gravity-mode regions. A smaller width means that the two regions are closer and hence stronger coupling exists between them. \cite{mos12a} developed an asymptotic relation for dipolar mixed modes based on the analysis of \cite{shiba79}:
\begin{equation}
 \nu = \nu_{n_p,\,l=1} + \frac{\Dnu_{\rm as}}{\pi}\arctan \left[q\tan \pi \left(\frac{1}{\Delta \Pi_1 \nu}-\epsilon_{\rm g} \right) \right],
 \label{eq:mixnu}
 \end{equation}
where $\nu_{n_p,\,l=1}$ are the uncoupled solutions for dipolar p modes, $q$ being the coupling strength and $\epsilon_{\rm g}$ is essentially the offset in equation~\eqref{eq:asyg}, which makes the obtained periods close to $(n+1/2+\epsilon_{\rm g})\Delta\Pi_l$ when the coupling is weak \citep{mos12a}. \cite{mos12a} used this expression to fit the observed frequencies that are extracted from the power spectra of \kepler\ red giants as well as clump stars, which gives the estimates of the coupling strength and period spacings. Equation~\eqref{eq:mixnu} provides an underlying link between observations and stellar inner structure, which was also discussed in some other studies \citep{mon13, jiang14,bos15,lag16}, and reveals how the mixed-mode frequency is shifted from the pure p-mode frequency as it couples with a pure g mode. The interaction between the p modes and the g modes takes place through a sequence of \textit{avoided crossings} \citep{osa75, aiz77}. At the avoided crossing the two modes with consecutive radial order approach very closely and exchange nature, but they actually do not cross and still preserve their original labelling. In this section, we discuss the mode coupling from a theoretical point of view by studying a pair of phase shifts that can help us understand the character of mixing and hence the mixed modes better.

\begin{figure}
%\epsscale{1.0}
\resizebox{1.0\hsize}{!}{\includegraphics{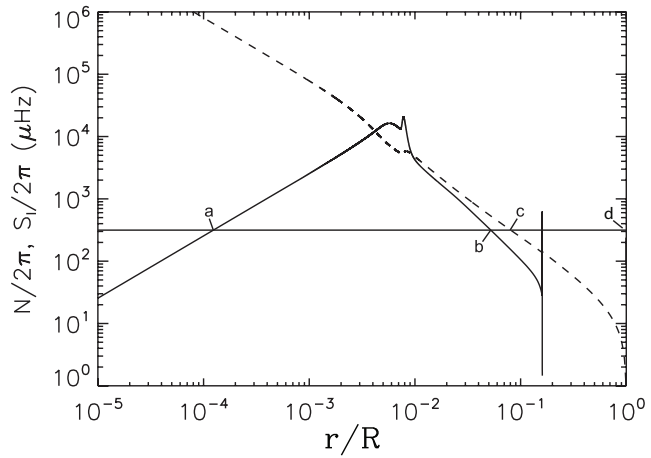}}
\caption{Propagation diagram of a RGB model $M_{\rm r}$. The buoyancy frequency (solid curve) and Lamb frequency (dashed curve, $l = 1$) are shown in terms of corresponding cyclic frequencies, against fractional radius $r/R$. The horizontal line indicates the mode frequency $\nu = 315.26 \, \muHz$, which presents strong mixed-mode character. The locations of the turning points of equation~\eqref{eq:k} are indicated by the letter $a$, $b$, $c$ and $d$, which divide the model into different propagation zones. $d$ is very close to the surface.}
\label{fg:sprop}
\end{figure}

\subsection{Eigenvalue conditions for mixed modes}
\cite{shiba79} initiated the analysis of eigenvalue conditions for different oscillatory modes using approximated solutions of Airy functions and presenting an eigenvalue condition for mixed modes with the coupling strength between gravity-wave and acoustic-wave cavities:
\begin{equation}
\cot\left(\int^{r_{\rm b}}_{r_{\rm a}}K {\rm d} r \right)\tan\left(\int^{r_{\rm d}}_{r_{\rm c}} K {\rm d} r \right)=\frac{1}{4}\exp\left(-2\int_{r_{\rm b}}^{r_{\rm c}} |K| {\rm d} r \right).
\label{eq:mix}
\end{equation}
Here the right-hand side defines the coupling strength $q$ and the wave number $K$ is approximated by
\begin{equation}
K^2 \approx \frac{\omega^2}{c^2}\left(1-\frac{S_{l}^{2}}{\omega^{2}}\right)\left(1-\frac{N^{2}}{\omega^{2}}\right).
\label{eq:k}
\end{equation}
$K$ is crucial in the asymptotic analysis, because the mode displacements have the property of being locally oscillating where $K > 0$. One the other hand, their properties switch to be locally exponential where $K < 0$. 
The interfaces between these regions are defined by the
{\it turning points} where $K = 0$.
It is obvious that the turning points confine propagation cavities of p and g modes \citep{unn89}. Therefore in equation~\eqref{eq:mix} $r_{\rm a}$, $r_{\rm b}$, $r_{\rm c}$ and $r_{\rm d}$ are the turning points, so the integral of the first term is computed within the gravity-mode cavity and that of the second term in the acoustic-mode cavity. The locations of the two cavities are illustrated by the propagation diagram, as shown for an RGB model ($M_{\rm r}$, introduced below) in Fig.~\ref{fg:sprop}. The frequency of a mixed mode is indicated by the horizontal line, which intersects the buoyancy frequency and the Lamb frequency at the four turning points labelled as $a$, $b$, $c$ and $d$. From the criteria of oscillation modes \citep{bookas}, we know that gravity-waves are trapped in the region between $a$ and $b$, in the core area, and acoustic-waves travel in the outer part of the star between $c$ and $d$. The evanescent region is therefore between $b$ and $c$. We note that this expression for $K$ places the upper acoustic turning point 
$r_{\rm d}$ immediately below the photospheric radius $R_{\rm ph}$. The limits of the integral interval for the coupling-strength term, on the right-hand side of equation~\eqref{eq:mix}, are the boundaries of the evanescent region. According to \cite{unn89} and equation~\eqref{eq:mix}, the coupling strength is assumed to be in the range $[0, 1/4]$. \cite{mos12a} found that $q$ can be larger than $1/4$ for clump stars but not beyond one when the coupling is at maximum. Indeed, \cite{tak16a, tak16b} found that when the evanescent region is thin, leading to strong coupling between the gravity waves in the core and the acoustic waves in the envelope, the expression for $q$ in equation~\eqref{eq:mix} should be modified, resulting in values that may approach one. A significant increase of $q$ is also observed in stars at the end of the subgiant phase where they transit to red giant phase~\citep{mos17}.

In this work we use equation~\eqref{eq:k} for $K$ in accordance with the analysis by \cite{shiba79} which we follow here.
The expression for $K$, however, depends on the variables that are used in the analysis.
An alternative expression, dealing more appropriately with the modes in the
near-surface region, was derived by Gough \citep{deu84,gou93} based on an analysis of a variable directly related to the Lagrangian pressure perturbation, following \cite{lam32}:
\begin{equation}
\mathcal{K}^2=\frac{\omega^2}{c^2}\left[1-\frac{\omega_{c}^{2}}{\omega^{2}}+\frac{S_{l}^{2}}{\omega^{2}}\left(\frac{N^{2}}{\omega^{2}}-1\right)\right],
\label{eq:kcutoff}
\end{equation}
with $\omega_{\rm c}$ being the cut-off frequency which is generally small in the stellar interior but can be large near the surface 
and it is neglected in equation~\eqref{eq:k}.
For high-order modes $\omega$ is comparable to $\omega_{\rm c}$ only at the photosphere where $\omega_{\rm c}$ is quite similar to $N^2$.
Hence the term with $\omega_{\rm c}^2$ in equation~\eqref{eq:kcutoff} can be replaced by $N^2$, such that we recover equation~\eqref{eq:k}.
However,
this approximation can result in large deviations for low-order p modes whose frequencies are well below $\omega_{\rm c}$ in the surface layers and for which $\omega_{\rm c}^2 / \omega^2$ is not negligible,
leading to a different eigenvalue condition \citep{gou93}. Therefore, for the mixed-mode eigenvalue condition in this paper we expect extra phase offsets appearing as deviations caused by the simplifications applied in the asymptotic analysis. We assume that the integral of $K$ may be corrected with a phase offset $\phi_{\rm p}$ for the p mode cavity. Therefore the eigenvalue condition becomes:
\begin{equation}
\cot\left(\int^{r_{\rm b}}_{r_{\rm a}} K {\rm d} r \right)\tan\left(\int^{r_{\rm d}}_{r_{\rm c}} K {\rm d} r - \phi_{\rm p} \right) = q .
\label{eq:fullcondi}
\end{equation}

In order to satisfy equation~\eqref{eq:fullcondi}, in the weak-coupling case where $q$ is small, a pair of conditions for $K$ is required:
 \begin{align}
 &  \int^{r_{\rm b}}_{r_{\rm a}}K {\rm d} r \approx \pi(n_{\rm g}+1/2) \pm \theta_{\rm g} \label{eq:gcon} \\
  & \int^{r_{\rm d}}_{r_{\rm c}} K {\rm d} r \approx \pi n_{\rm p} \mp \theta_{\rm p} + \phi_{\rm p} \;, \label{eq:pcon_cor}
\end{align}
where integers $n_{\rm g}$ and $n_{\rm p}$ are the radial orders of uncoupled g and p modes, respectively. $\theta_{\rm g}$ and $\theta_{\rm p}$ are offsets for g and p mode, each of the order of magnitude of  $\pi \sqrt{q}$ \citep{shiba79}, and they appear as phase shifts in the trigonometric functions in equation~\eqref{eq:mix}. The physical meaning of $\theta_{\rm g}$ and $\theta_{\rm p}$ is clear: they represent the deviations of modes undergoing avoided crossing from pure modes. As a consequence of mode coupling, determined by $q$ in equation~\eqref{eq:mixnu}, the frequencies of mixed modes vary around that of the pure mode. This wobbling character of the modes is implied by the very term $\theta_{\rm g}$ and $\theta_{\rm p}$ and the $\pm$ and $\mp$ signs. 
In the case of no coupling, with $q = 0$, $\theta_{\rm g}$ and $\theta_{\rm p}$ are also $0$ and equations~\eqref{eq:gcon} and \eqref{eq:pcon_cor} reduce to the asymptotic relations for the pure g and p modes, respectively. For simplicity, the $\pm$ signs in equations~\eqref{eq:gcon} and \eqref{eq:pcon_cor} are not used hereafter and they are included as the signs of the values of the $\theta$s. The estimation of $\phi_{\rm p}$ is discussed in detail in Section~\ref{sc:phi_p}.

\begin{table}
\caption{Basic properties of Model $M_{\rm s}$ and $M_{\rm r}$.}
\label{tb:models}
\center
\scalebox{0.85}{
\begin{tabular}{@{}cccccccc}
\hline
& {\large $M$} &{\large Age} & {\large $T_{\rm eff}$} & {\large $L$} & \large $R$ & {\large $\Dnu_{\rm obs}$} & \large {$\Delta\Pi_{\rm obs}$} \\
&\small {(M$_{\sun}$)} &\small{(Gyr)} & \small{(K)} & \small{(L$_{\sun}$)} & \small{(R$_{\sun}$)} & \small{($\muHz$)} & \small{(s)}\\
\hline
\small{$M_{\rm s}$} & \small 1.3 & \small 4.34 & \small 5835.6 & \small 4.887 & \small 2.168 & \small 49.28 & \small 350 \\ 
\small{$M_{\rm r}$}  & \small 1.3 & \small 4.90 & \small 4772.2 & \small 8.191 & \small 4.198 & \small 17.81 & \small 87.27 \\
\hline
\end{tabular}
}
\end{table}

It is of great importance to understand $\theta_{\rm p}$ and $\theta_{\rm g}$ in the study of mixed modes. To verify the theory about these phase shifts we computed a 1.3 M$_{\sun}$ model with initial hydrogen abundance $X = 0.7$ and heavy element abundance $Z = 0.02$, using the {\scriptsize ASTEC} evolution code \citep{jcd08a} and the corresponding oscillation frequencies with the {\scriptsize ADIPLS} oscillation package \citep{jcd08b}. Convection is treated under the assumption of mixing-length theory \citep{boh58} with the mixing-length parameter set to 1.8. The input physics of the current modelling includes the latest OPAL opacity tables \citep{igl96}, OPAL equation of state in its 2005 version \citep{rog96}, and NACRE reaction rates \citep{ang99}. The models are kept as simple as possible, meaning that there is no overshooting, no diffusion, and no rotation in the modelling.
In order to make the analysis easy, the fourth-order system of differential equations is reduced to second order using the Cowling approximation \citep{cow41}, neglecting the Eulerian perturbation to the gravitational potential%
\footnote{However, we note that for dipolar modes it is possible to reduce the fourth-order system to second order without neglecting the perturbation to the gravitational potential \citep{tak06}.}.
Along the evolutionary track, one SGB model, $M_{\rm s}$, is selected to illustrate the case of p modes undergoing avoided crossings, and one RGB model, $M_{\rm r}$, is selected to illustrate the case where g modes are dominant. 
Some basic information on model $M_{\rm s}$ and $M_{\rm r}$ can be found in Table~\ref{tb:models}.
Similar to what is found for the MS, on the SGB high-order p modes dominate the power spectrum. However, they are mixed with g modes to different degrees that relate to $\theta_{\rm p}$. On the other hand, at a more evolved evolutionary stage, the oscillation power spectrum is rich in g modes, some of which can be detected in terms of mixed modes that have significant p-mode character. 
We will follow the classification of mixed modes introduced by \citet{mos12b} and \citet{goup13} that divides them into two categories, p-m and g-m mixed modes; this is determined by the location of the dominant contribution to the kinetic energy \citep{jiang14}. If a mixed mode has a larger contribution from the acoustic cavity in mode energy than its neighbour, it behaves more like a p mode and hence it is called p-m mixed mode, otherwise it is a g-m mixed mode. 
We note that p-m modes have larger amplitudes at the surface, making them generally more easily detectable than the g-m modes \citep{dup09, gro14}.
We emphasis the importance of taking $\phi_{\rm p}$ into consideration when the mixed modes eigenvalue conditions are used in a global analysis as in the examples given in Sections~\ref{sc:mc} and~\ref{sc:order}, after the introduction of estimating $\phi_{\rm p}$ theoretically using the eigenfunctions in next section.

\begin{table}
\caption{Radial orders for Model $M_{\rm s}$ dipolar mixed modes. $n_{\rm p}$ and $n_{\rm g}$ are the radial orders of the p- and g-mode components that contribute to the corresponding mixed mode, from equation~\eqref{eq:corthetap} and \eqref{eq:eg}. The numerical radial orders are notated as $\mathcal{N}_{\rm p}$ and $\mathcal{N}_{\rm g}$ that gives the mode radial order $n$. $\phi_{\rm p}$ is obtained by fitting the eigenfunction $\xi_{\rm r}$, to correct the deviations due to the approximations in the asymptotic analysis. The mode frequencies are given in the sixth column.}
\label{tb:msp}
\center
\begin{tabular}{@{}cccccrrrr}
\hline
\small{$n$} & \small{$n_{\rm p}$}& \small{$n_{\rm g}$} &\small{$\mathcal{N}_{\rm p}$} &\small{$\mathcal{N}_{\rm g}$} & \small{$f (\muHz)$} & \small{$\theta_{\rm p} / \pi$} & \small{$\theta_{\rm g} / \pi$} & \small{$\phi_{\rm p} / \pi$} \\
\hline
\small 1 & \small 6 & \small 5  &\small 6 &\small 5 & \small 406.12 &\small 0.28 &\small 0.03 &\small 1.03 \\
\small 2 & \small 7 & \small 5 &\small 7 &\small 5 & \small 440.55 &\small $-0.05$ &\small 0.50 &\small 1.01 \\
\small 3 & \small 7 & \small 4 &\small 7 &\small 4 & \small 462.91 &\small 0.40 &\small 0.18 &\small 0.98 \\
\small 4 & \small 8 & \small 4 &\small 8 &\small 4 & \small 492.06 &\small $0.01$ &\small $-0.17$ &\small 0.93 \\
\small 5 & \small 9 &  \small 3 &\small 9 &\small 4 & \small 536.51 &\small $-0.05$ &\small 0.36 &\small 0.83 \\
\small 6 & \small 9 & \small 3 &\small 9 &\small 3 & \small 560.49 &\small 0.45 &\small 0.15 &\small 0.78 \\
\small 7 & \small 10 & \small 3 &\small 10 &\small 3 & \small 585.53 &\small $-0.01$ &\small $-0.06$ &\small 0.72 \\
\small 8 & \small 11 & \small 3 &\small 11&\small 3 & \small 632.90 &\small $-0.04$ &\small $-0.40$ &\small 0.65 \\
\small 9 & \small 12 & \small 2 &\small 12 &\small 3 & \small  681.05 &\small $-0.06$ &\small 0.30 &\small 0.59 \\
\small 10 & \small 13 & \small 2 &\small 13 &\small 3 & \small 725.95 &\small -0.13 &\small 0.05 &\small 0.51 \\
\small 11 & \small 13 &  \small 2 &\small 13 &\small 2 & \small 739.83 &\small $0.17$ &\small $-0.01$ &\small 0.48 \\
\small 12 & \small 14 & \small 2 &\small 14 &\small 2 & \small 778.29 &\small -0.04 &\small $-0.20$ &\small 0.42 \\
\small 13 & \small 15 & \small 2 &\small 15 &\small 2 & \small 826.39 &\small -0.05 &\small $-0.39$ &\small 0.35 \\
\small 14 & \small 16 & \small 1 &\small 16 &\small 2 & \small 875.66 &\small -0.06 &\small 0.43 &\small 0.30 \\
\small 15 & \small 17 & \small 1 &\small 17 &\small 2 & \small 925.08 &\small -0.08 &\small 0.27 &\small 0.27 \\
\small 16 & \small 18 & \small 1 &\small 18 &\small 2 & \small 974.04 &\small -0.09 &\small 0.13 &\small 0.22 \\
\small 17 & \small 19 & \small 1 &\small 19 &\small 2 & \small 1021.74 &\small -0.14 &\small 0.01 &\small 0.18 \\
\small 18 & \small 20 & \small 1 &\small 19 &\small 1 & \small 1058.19  &\small $-0.41$ &\small $-0.08$ &\small 0.15 \\
\small 19 & \small 20 & \small 1 &\small 20  &\small 1 & \small 1082.39 &\small 0.09 &\small $-0.14$ &\small 0.12 \\
\small 20 & \small 21 &  \small 1 &\small 21 &\small 1& \small 1127.27 &\small -0.03 &\small $-0.23$ &\small 0.12 \\
\small 21 & \small 22 &  \small 1 &\small 22 &\small 1 & \small 1176.34 &\small -0.06 &\small $-0.33$ &\small 0.12 \\
\small 22 & \small 23 &  \small 1 &\small 23 &\small 1 & \small 1226.04 &\small -0.07 &\small $-0.41$ &\small 0.11 \\
\small 23 & \small 24 &  \small 1 &\small 24 &\small 1 & \small 1275.86 &\small -0.08 &\small $-0.49$ &\small 0.10 \\
\small 24 & \small 25 & \small 0 &\small 25 &\small 1 & \small 1325.57 &\small $-0.09$ &\small 0.43 &\small 0.08 \\
\small 25 & \small 26 & \small 0 &\small 26 &\small 1 & \small 1375.05 &\small $-0.11$ &\small 0.36 &\small 0.06 \\
\hline
\end{tabular}
\end{table}

\begin{figure}
\resizebox{1.0\hsize}{!}{\includegraphics{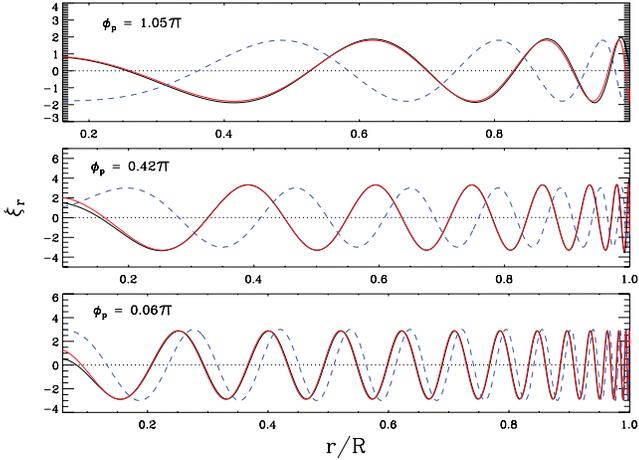}}
\caption{The scaled radial displacement of three $M_{\rm s}$ dipolar mixed modes, $n = 1, 13, 26$ from top to bottom, in the p-mode cavity as a function of the radius fraction. The turning points of each region are located at the edges of each plot. The scaled $\xi_{\rm r}$ from models is in black and that from asymptotic estimations tuned with the phase offset $\phi_{\rm p}$ (given in each plot) is in red while those without $\phi_{\rm p}$ are the blue dashed lines.}
\label{fg:dis}
\end{figure}

\subsection{Eigenfunctions and $\phi_{\rm p}$} \label{sc:phi_p}
The eigenvalue condition in equation~\eqref{eq:fullcondi} shows that the coupling of mixed modes is related to the integral of $K$ that provides us a useful tool to explore the interior structure of the star. The integral can be gained from models as the essential characteristic frequencies ($N^2$ and $S^2_{l}$) and the sound speed are computed during the modelling. Before discussing the character of mixed modes, we present how to estimate $\phi_{\rm p}$ using the eigenfunctions. Since the numerical radial order $\mathcal{N}_{\rm p}$ and $\mathcal{N}_{\rm g}$ computed in {\scriptsize ADIPLS} is given by the number of nodes in the eigenfunction in the radial direction, it is interesting to examine the differences between the asymptotic and numerical eigenfunctions, i.e. the oscillatory radial displacement $\xi_{\rm r}$. In Fig.~\ref{fg:dis} the scaled $\xi_{\rm r}$ is plotted in the p-mode cavity for three p modes in model $M_{\rm s}$. The $\xi_{\rm r}$ is computed and scaled to get a constant amplitude according to their asymptotic behaviour \citep[][Appendix E.2]{bookas}, assuming $S_l^2 \ll \omega^2$ in the p-mode cavity:
\begin{equation}
\xi_{\rm r} \simeq A(\rho \omega c)^{-1/2} r^{-1} \cos \left(\int_{r}^{r_{\rm d}} K {\rm d} r - (\alpha+ \frac{1}{4}) \pi \right),
\label{eq:xir}
\end{equation}
where $A$ is a constant and $\alpha$ is a phase that depends on the structure of the surface layers (cf. section 5 of Gough 1993 and appendix E of Aerts et al. 2010). Based on the Duvall law \citep{duv82}, $\alpha$ can be estimated according to Appendix E.3 of Aerts et al. 2010. However, $\alpha$ is essentially approximated by our $\phi_{\rm p}$ in p mode region because near the turning point $r_{\rm c}$ the integral term in equation~\eqref{eq:xir} is approximated by that in the eigenvalue condition of equation~\eqref{eq:fullcondi}. Therefore in this way $\phi_{\rm p}$ is estimated by matching up the asymptotic $\xi_{\rm r}$ near $r_{\rm c}$ with the numerical one. The resulting $\phi_{\rm p}$ of each $M_{\rm s}$ mode is given in the last column of Table~\ref{tb:msp}. The values of $\phi_{\rm p}$ are around one for low-order modes for which $\omega_{\rm c}$  is rather different from $N$ at the upper turning point of the mode. They diminish to close to 0 for high-order modes whose frequencies are comparable to $\omega_{\rm c}$ only where $\omega_{\rm c}$ is similar to $N$, making the definition of $K$ given by equation~\eqref{eq:k} a good approximation of equation~\eqref{eq:kcutoff}. The eigenfunction is one way to examine $\phi_{\rm p}$ as well as mixed modes when we know the interior property of stars which is normally not possible from only observations. In Section~\ref{sc:obs}, a different way to estimate $\phi_{\rm p}$ linking to the observed seismic parameters is discussed.

\begin{figure}
%\epsscale{1}
\resizebox{1.0\hsize}{!}{\includegraphics{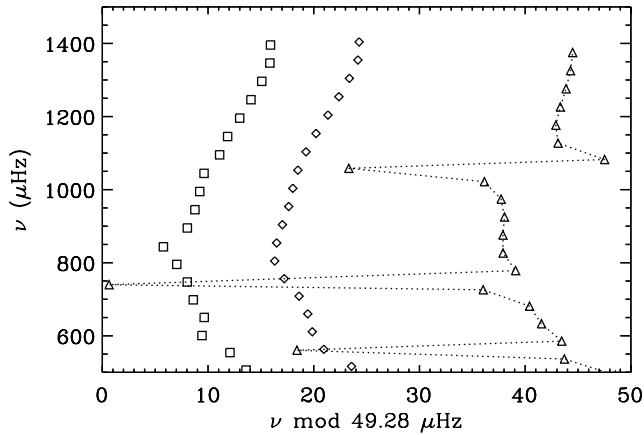}}
\caption{Frequency \'{e}chelle diagram for modes in $M_{\rm s}$. Modes of different degrees are indicated by diamonds ($l = 0$), triangles ($l = 1$) and squares ($l = 2$). To guide the eyes the dashed line connects the dipolar mixed modes. Quadrupolar ($l = 2$) g-m mixed modes are suppressed in the plot.}
\label{fg:msp}
\end{figure}

\subsection{Mode coupling} \label{sc:mc}

A mixed mode simultaneously presents characters of oscillations residing in gravity-wave and acoustic-wave cavities. The width of the evanescent region depends on mode frequency and the two characteristic frequencies $N$ and $S_l$ and, therefore, on the interior structure. From previous studies \citep{ben12,mos12a,jiang14}, we know that $q$ is a global measurement of the coupling. The variation of the coupling strength in a given star is very small over the relevant frequency range in the weak-coupling case~\citep{mos17}, but the mixed character of each mode differs to different degrees. Therefore the coupling strength factor $q$ is not the parameter that optimally characterizes the behaviour of mixing of individual mixed modes. From equations~\eqref{eq:gcon} and~\eqref{eq:pcon_cor} we can anticipate that $\theta_{\rm g}$ and $\theta_{\rm p}$ are the parameters that really matter in that respect.

The frequency \'{e}chelle diagram of oscillation modes of model $M_{\rm s}$  is shown in Fig.~\ref{fg:msp}, for clarity suppressing the g-m modes with $l = 2$. Though there are slight shifts, the radial modes (diamonds) are stacked up vertically, because they are nearly equally spaced by the large frequency separation. Mixed modes for $l = 1$ are clearly illustrated in Fig.~\ref{fg:msp} as their frequencies deviate substantially from the vertical pattern. This is due to the fact that when an avoided crossing happens the mode frequency bumps to a higher value for a g-m mixed mode. 
%As mentioned above, $q$ does not vary much from mode to mode and hence it does not reflect the occurrence of avoided crossing explicitly.
The degree of mixing for each mode is decided by the phase shift; following equation~\eqref{eq:pcon_cor} for the p mode case it is estimated by:
%\begin{equation}
%\theta_{\rm p} \approx \int_{r_{\rm c}}^{r_{\rm d}} K {\rm d} r - \pi n_{\rm p}.
%\label{eq:ep}
%\end{equation}
%Since $K$ depends on mode frequency and $l$ for a given model, $\theta_{\rm p}$ also varies with the frequency, if we only consider dipolar modes. A more ideal expression of $\theta_{\rm p}$ is  retrieved by taking $\phi_{\rm p}$ into the equation:
 \begin{equation}
   \theta_{\rm p} = \int_{r_{\rm c}}^{r_{\rm d}} K {\rm d} r - \pi n_{\rm p} - \phi_{\rm p} .
    \label{eq:corthetap} 
\end{equation}

\begin{figure}
\resizebox{1.0\hsize}{!}{\includegraphics{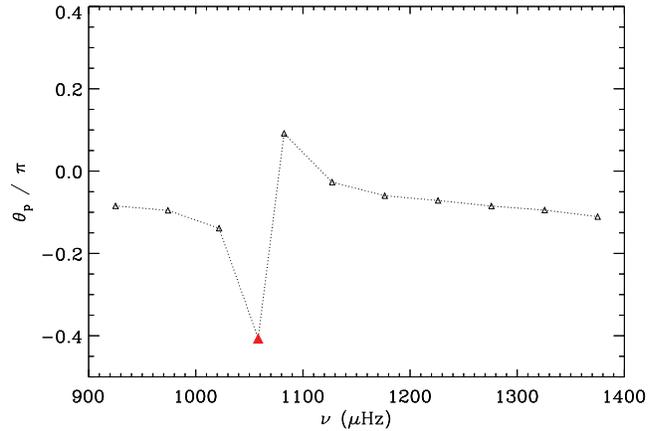}}
\caption{Computed $\theta_{\rm p}$ of dipolar modes of Model $M_{\rm s}$ from equation~\eqref{eq:corthetap} as a function of mode frequency. Each mode is illustrated by a triangle. The mode undergoing the most significant mixing, the most g-mode-like mixed mode, is filled in red.}
\label{fg:esp}
\end{figure}

Fig.~\ref{fg:esp} shows the resulting $\theta_{\rm p}$ of $l = 1$ modes for Model $M_{\rm s}$, using the $\phi_{\rm p}$ obtained by fitting the eigenfunctions. Only modes in the high-frequency range are plotted in the figure. The mode undergoing significant avoided crossing, or mixing, and having a substantial jump in $\theta_{\rm p}$  compared to normal p modes has a frequency around 1030 $\muHz$ and its $\theta_{\rm p}$ is very close to $-0.5\pi$. The $\theta_{\rm p}$ of regular p modes is very small (around 0) indicating that they preserve more p-mode character. The large jump in $\theta_{\rm p}$ of the g-m mixed mode is so obvious in this SGB model that the use of $\theta_{\rm p}$ as an indicator of the degree of mixing is promising. 

\begin{figure}
\resizebox{1.0\hsize}{!}{\includegraphics{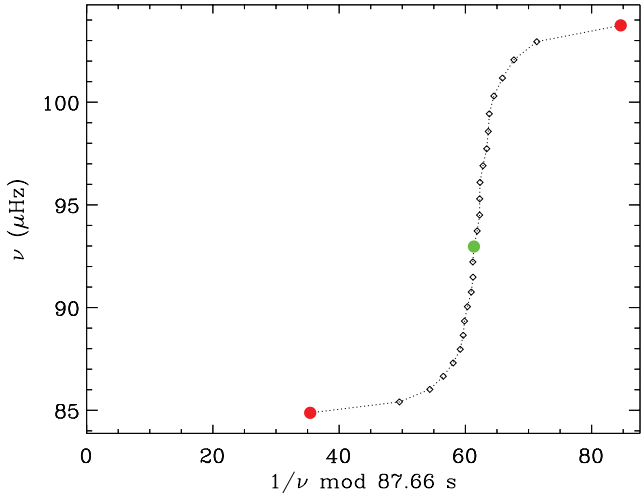}}
  \caption{Period \'{e}chelle diagram of a segment of gravity modes in $M_{\rm r}$. These g modes are coupled with two p mode. The modes in red are the most p-mode-like mixed modes among all the modes in this plot, while the green circle indicates the one that preserve most g-mode character during the coupling.}
  \label{fg:mrg}
\end{figure}

\begin{figure}
\resizebox{1.0\hsize}{!}{\includegraphics{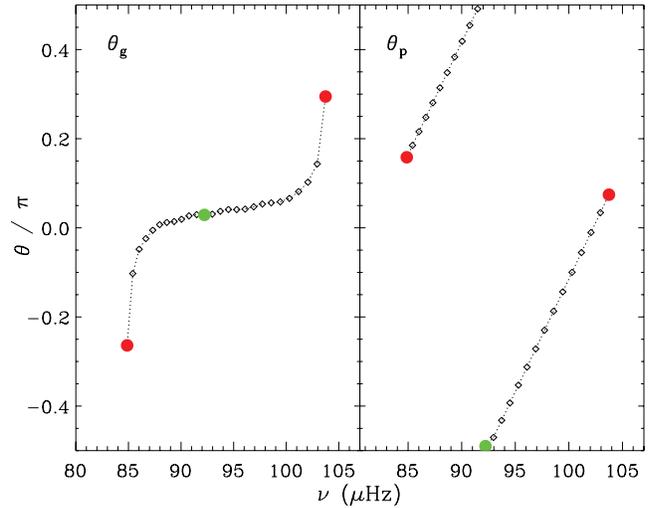}}
\caption{Left: Computed $\theta_{\rm g}$ of dipolar modes of Model $M_{\rm r}$ from equation~\eqref{eq:eg} as a function of frequency; Right: $\theta_{\rm p}$ computed from equation~\eqref{eq:corthetap}. Green and red circles correspond to the same kinds of mixed modes as in Fig.~\ref{fg:mrg}.}
\label{fg:erg}
\end{figure}

In the case of RGB models in which modes are much denser in frequency, a few p-m mixed modes could be observed for each avoided crossing and $\theta_{\rm g}$ and $\theta_{\rm p}$ vary as the modes experience different degrees of mixing. As shown in Fig.~\ref{fg:mrg}, a period \'{e}chelle diagram for the model $M_{\rm r}$ is produced by plotting the frequencies as a function of the period modulo the period spacing \citep{bed11,mos12a}. The uncoupled g modes should be aligned vertically in this diagram, but in practice modes spread sideward due to the effect of coupling, leaving the most p-m mixed modes (red circles in the figure) located at the sides of the s-shape pattern and the most g-m one (green circle) located at the centre. In Fig.~\ref{fg:mrg}, all these modes are coupled between different orders of g modes and two p modes with p-mode radial order $n_{\rm p}$ = 3 and 4. Modes with frequencies larger than the central g-m mode couple with the $n_{\rm p} = 4$ p mode while the rest couple with the  $n_{\rm p} = 3$ p mode. The mixing degrees of these mixed modes are linked to the $\theta_{\rm g}$ of each mode. The shift term $\theta_{\rm g}$ is now given by:
\begin{equation}
\theta_{\rm g} \approx \int_{r_{\rm a}}^{r_{\rm b}} K {\rm d} r - \frac{\pi}{2} - \pi n_{\rm g} .
\label{eq:eg}
\end{equation}
The values of $\theta_{\rm g}$ for the corresponding mixed modes are illustrated in Fig.~\ref{fg:erg}(a). Modes in red are those having most p-mode-like character and located at the edges of the pattern in Fig.~\ref{fg:mrg}. Their absolute values of $\theta_{\rm g}$ (near $0.5 \pi$) are much larger than those of less perturbed g-mode-like modes (the green mode and its neighbours in Fig.~\ref{fg:mrg}), meaning they are the closest to a pure p mode while the green mode is the closest to a pure g mode. As the absolute value of $\theta_{\rm g}$ approaches $0.5 \pi$, the modes behave more like a p mode. Meanwhile, in Fig.~\ref{fg:erg}(b) $\theta_{\rm p}$ is computed from equation~\eqref{eq:corthetap}. Here the modes marked in red have $\theta_{\rm p}$ very close to 0 indicating the purity of their p-mode character. Similarly, the mode marked in green, the most g-mode-like one, has less p-mode character than any other mixed modes in the plot and therefore an absolute $\theta_{\rm p}$ value close to $0.5 \pi$ is associated with it. 

%The phase shifts $\theta_{\rm g}$ and $\theta_{\rm p}$ are different from the $\epsilon_{\rm p}$ and $\epsilon_{\rm g}$ that appear in the frequency and period asymptotic pattern. \cite{mos15} defined the two phases including $\epsilon_{\rm p}$ and $\epsilon_{\rm g}$ and using the spacings $\Dnu_{\rm as}$ and $\Delta\Pi$. In Section~\ref{sc:obs} we will briefly discuss the link between $\epsilon_{\rm p}$ and $\phi_{\rm p}$.
In most cases the comparisons between the mixed-mode frequencies and the expected pure p-mode frequency can help estimate how much a mode deviates from a pure p or g mode, except when the structural glitches in the cores of red giants significantly influence the inertia and frequencies of their mixed modes \citep{cun15}. 
The $\theta_{\rm g}$ and $\theta_{\rm p}$ provides an alternative way to measure the degree of mixing in glitch-free cases and their measurement is related to the integrals over different regions. So the method presented here is different from that which was used by \cite{whi12} for their mode identification in F stars. They measured $\epsilon_{\rm p}$, assumed to be independent of frequency, from the frequencies and several observed seismic parameters.  
Here we consider the phases $\theta_{\rm g}$ and $\theta_{\rm p}$ as indicators of the offset from the pure asymptotic relations for mixed modes, and hence as a window to study the mode variations caused by coupling.
Further information about the properties of a mixed mode can be obtained from
the ratio between the contributions of the p and g cavities to the kinetic energy.
\cite{ben14} demonstrated a way to estimate the mode-inertia variations between neighbouring modes from the observed mode amplitudes and line widths
which provide information about the mode coupling and a potential 
diagnostics of the stellar properties.
However, here we focus on the use of frequencies as diagnostics, noting that the integrals of the wave number $K$ are related to two observable parameters, namely, $\Delta\Pi$ and $\Dnu_{\rm as}$, which makes an analysis based on the phase shifts possible. This will be discussed below.

\subsection{Radial order of mixed modes} \label{sc:order}

With the help of models we can study properties of mixed modes by deriving some parameters that characterize oscillation modes, such as kinetic energy \citep{unn89} or mode inertia \citep{dzi01}. The normalised mode inertia is given by:
\begin{equation}
E=\frac{\int_0^R\left(\xi_{\rm r}^2+L^2\xi_{\rm h}^2\right)\rho r^2{\rm d} r}{M \left[ \xi_{\rm r}(R_{\rm ph})^2 + L^2 \xi_{\rm h} (R_{\rm ph})^2 \right]} ,
\end{equation} 
where $\rho$ is the density and $\xi_{\rm h}$ is the oscillatory displacement in the horizontal direction. For a mode that is dominated by p-mode nature, the greatest contribution to the inertia comes from the p-mode cavity, hence from $\xi_{\rm r}$. For a mixed mode that behaves predominantly as a g mode, the inertia is very large owing to the high density in the core area and hence is dominated by $\xi_{\rm h}$ in the g-mode cavity \citep{bookas}.  Therefore, a convenient way to measure the g nature of mixed modes is through the ratio of mode inertia in the g cavity over the total inertia:
\begin{equation}
\zeta=\frac{E_{\rm g}}{E} \approx \frac{\int_{r_{\rm a}}^{r_{\rm b}} L^2\xi_{\rm h}^2\rho r^2{\rm d} r}{\int_0^R\left(\xi_{\rm r}^2+L^2\xi_{\rm h}^2\right)\rho r^2{\rm d} r} .
\end{equation}
\cite{goup13} have originally introduced the parameter $\zeta$ and made use of it to reproduce the rotation splittings by comparing the radial and dipolar mode inertia. Later on \cite{deh14} have proposed a more precise expression of $\zeta$ for the aim of probing the radial dependence of the rotation profiles for a sample of \kepler\ targets. \cite{mos15} showed that the bumped period spacing and frequency rotational splitting follow the same pattern constructed by $\zeta$. The value of $\zeta$ is close to 0 for a p-dominated mode and close to 1 for a g-dominated one, making $\zeta$ an optimal parameter to identify the type of mixed modes;
%However, the expression of $\zeta$ includes the coupling strength $q$ \citep[equation~(14) in][]{mos15} that is not easily observable.
on this basis \cite{mos17} developed an automated method to estimate the coupling strength $q$ and applied it to a large number of stars observed by \kepler .
Here we investigate the p and g contributions to the mixed modes by measuring $n_{\rm p}$ and $n_{\rm g}$ through $\theta_{\rm p}$ and $\theta_{\rm g}$ which do not require to compute $q$.

\begin{figure}
\resizebox{1.0\hsize}{!}{\includegraphics{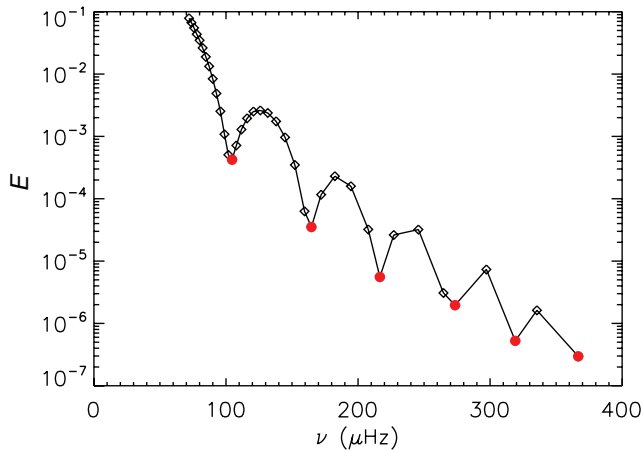}}
\caption{Normalised mode inertia of $l = 1$ g modes ($n < 0$) for the model $M_{\rm s}$. 
The p-m modes are plotted in red, indicating that the six lowest-order p modes couple with g modes in this frequency range.}
\label{fg:inertia}
\end{figure}

In equations~\eqref{eq:gcon} and \eqref{eq:pcon_cor}, $n_{\rm p}$ and $n_{\rm g}$ represent the radial orders for what would be the pure p- and g-mode components contributing to a given mixed mode. Therefore, by using the two equations we here introduce a new way to evaluate $n_{\rm p}$ and $n_{\rm g}$ as well as the resulting radial order $n$ for a mixed mode. Since only the integral terms are directly obtainable, the phase shift and radial orders in equations~\eqref{eq:gcon} and \eqref{eq:pcon_cor} are essentially considered together. From Section~\ref{sc:mc} we know that $\theta_{\rm p}$ and $\theta_{\rm g}$ of a mixed mode vary between $-0.5 \pi$ and $0.5 \pi$, and hence $n_{\rm p}$ and $n_{\rm g}$ can be estimated by rounding the values of $\pi^{-1} \left( \int_{r_{\rm c}}^{r_{\rm d}} K {\rm d} r - \phi_{\rm p} \right)$ and $\pi^{-1} \int_{r_{\rm a}}^{r_{\rm b}} K {\rm d} r - \frac{1}{2}$, respectively.

We first use the model $M_{\rm s}$ to discuss the shift in p-mode radial order caused by the coupling. Fig.~\ref{fg:msp} shows p modes that experience coupling with several g modes, although the coupling already happens in the low frequency range where g modes exist. The normalised mode inertia $E$ of g modes in Model $M_{\rm s}$, plotted in Fig.~\ref{fg:inertia}, is large because of their large horizontal displacements in the deep interior where the density is high. The add-in of p-mode character increases the displacement vertically in the outer part and suppresses it horizontally, leading to a decrease in inertia. As a result, those mixed modes having locally small inertia, which are the red ones and their neighbours in the plot, are p-m mixed modes and are more likely to be detected in observations.

\begin{table}
\caption{Similar to Table~\ref{tb:msp}, but the radial orders for Model $M_{\rm r}$ dipolar mixed modes. Only a few high order p-m (odd row) and g-m (even row) mixed modes with frequencies around $\numax$ are listed. The $\theta_{\rm p}$, using $\phi_{\rm p}$, and corresponding $n_{\rm p}$ are given here. $\phi^{\prime}_{\rm p}$ is estimated using equation~\eqref{eq:phi_final}.}
\label{tb:msg}
\center
\scalebox{0.8}{
\begin{tabular}{@{}cccccrrrrrr}
\hline
\small{$n$} & \small{$n_{\rm p}$} & \small{$n_{\rm g}$} &\small{$\mathcal{N}_{\rm p}$} &\small{$\mathcal{N}_{\rm g}$}  & \small{$f (\muHz)$} & \small{$\theta_{\rm p} / \pi$} & \small{$\theta_{\rm g} / \pi$} & \small{$\phi_{\rm p} / \pi$}  & \small{$\phi^{\prime}_{\rm p} / \pi$}   \\
\hline
\small $-49$ & \small 9 & \small 58 &\small 9 &\small 58 & \small 192.24 & \small $-0.04$ & \small $-0.34$ & \small 0.68 & \small 0.66 \\
\small $-46$ & \small 9 & \small 55 &\small 9 &\small 55 & \small 201.11 & \small $0.49$ & \small 0.04 & \small 0.60 \\
\small $-43$ & \small 10 & \small 53 &\small 10 &\small 53 & \small 209.64  & \small $-0.01$ & \small $-0.33$ & \small 0.55 & \small 0.55 \\
\small $-41$ & \small 10 & \small 51 &\small 10 &\small 51 & \small 216.41  & \small 0.38 & \small 0.03 & \small 0.52 \\
\small $-38$ & \small 11 & \small 49 &\small 11 &\small 49 & \small 226.99  & \small $-0.04$ & \small $-0.43$ & \small 0.49 & \small 0.47\\
\small $-36$ & \small 11 & \small 47 &\small 11 &\small 47 & \small 234.18  & \small 0.38 & \small 0.02 & \small 0.45 \\
\small $-33$ & \small 12 & \small 45 &\small 12 &\small 45 & \small 245.29  & \small $0.01$ & \small $-0.28$ & \small 0.40& \small 0.40 \\
\small $-31$ & \small 12 & \small 43 &\small 12 &\small 43 & \small 254.99  & \small 0.49 & \small 0.04 & \small 0.36 \\
\small $-29$ & \small 13 & \small 42 &\small 13 &\small 42 & \small 262.74  & \small $0.01$ & \small $-0.28$ & \small 0.32 & \small 0.32 \\
\small $-27$ & \small 13 & \small 40 &\small 13 &\small 40 & \small 273.23  & \small 0.48 & \small 0.06 & \small 0.30  \\
\small $-25$ & \small 14 & \small 39 &\small 14 &\small 39 & \small 281.30  & \small $0.04$ & \small $-0.14$ & \small 0.27 & \small 0.26 \\
\small $-24$ & \small 14 & \small 38 &\small 14 &\small 38 & \small 287.22   & \small 0.37 & \small 0.02 & \small 0.25 \\
\small $-22$ & \small 15 & \small 37 &\small 15 &\small 37 & \small 298.37  & \small $-0.02$ & \small $-0.46$ & \small 0.24 & \small 0.20 \\
\small $-20$ & \small 15 & \small 35 &\small 15 &\small 35 & \small 310.34  & \small 0.49 & \small 0.06 & \small 0.23 \\
\hline
\end{tabular}
}
\end{table}

The red modes in Fig.~\ref{fg:inertia} indicate that the six lowest-order p modes couple with g modes and their frequencies are reduced to lower values making them buried in the g-mode frequency region. Table~\ref{tb:msp} lists the radial orders of the components concealed in each mixed mode in model $M_{\rm s}$, namely the radial order $n_{\rm p}$ and $n_{\rm g}$ of uncoupled p- and g-mode components. We have to note that the orders are in integer numbers only because they are rounded from the values of the integral terms. As a comparison, in the table we also provide the orders $\mathcal{N}_{\rm p}$ and $\mathcal{N}_{\rm g}$ that are calculated by {\scriptsize ADIPLS} which applies the mode classification scheme introduced in Chapter 17 of \cite{unn89}. The resulting radial order $n$ of a mixed modes is then given as $\mathcal{N}_{\rm p} - \mathcal{N}_{\rm g}$. In Table~\ref{tb:msp} the $\theta_{\rm p}$ and $n_{\rm p}$ are computed using equation~\eqref{eq:corthetap}, where the $\phi_{\rm p}$ values are estimated in Section~\ref{sc:phi_p}. After taking $\phi_{\rm p}$ into consideration we found identical $n_{\rm p}$ and $\mathcal{N}_{\rm p}$ values for all modes with the exception of the $n = 18$ mode%
\footnote{This is due to our way of estimating $n_{\rm p}$, increasing it by 1 when $\theta_{\rm p}$ reaches $0.5 \pi$. Allowing a slightly larger $\theta_{\rm p}$ in absolute magnitude would have yielded the `correct' $n_{\rm p}$.} (the red mode in Fig.~\ref{fg:esp}) which is undergoing strong mixing and has a $\theta_{\rm p}$ close to $-0.5 \pi$. Given the fact that the $\phi_{\rm p}$ is approximated by arbitrarily tuning the asymptotic $\xi_{\rm r}$, the determination of the $n_{\rm p}$ becomes a little tricky for those g-m mixed modes with $|\theta_{\rm p}|$ around $0.5 \pi$. The differences in g-mode orders for the four p-m mixed modes, with $n$ around 5, 10, 17 and 25 in Table~\ref{tb:msp}, indicate that there is possibly an additional frequency-dependent phase for $\theta_{\rm g}$ for SGB models of which the high-order modes can be strong-coupling modes.

%\textbf{The differences in g-mode orders for the four p-m mixed modes, with $n$ around 5, 10, 17 and 25 in Table~\ref{tb:msp}, are more possibly caused by the fact that the high-order modes in the SGB model are strong-coupling modes of which $q$ is not a small value and may be over $0.25$. Therefore, according to equation~\eqref{eq:mix} an additional phase should appear on the right-hand side of equation~\eqref{eq:eg} for $\theta_{\rm g}$ while equation~\eqref{eq:corthetap} almost remains the same.}

 A more illustrative example is given by the p-m mixed modes in the RGB model. In Table~\ref{tb:msg} we present the radial orders for the 7 most p-m like mixed modes with consecutive $n_{\rm p}$ (odd rows) and the 7 most g-m like mixed modes (even rows) for the RGB model $M_{\rm r}$. These modes have frequencies around $\numax$%
\footnote{For models we estimate $\numax$ using the scaling relation given by equation (10) in \cite{kje95}, in terms of surface gravity and effective temperature.},
i.e., the frequency of the maximum oscillation power, as do the detectable RGB mixed modes. The asymptotic orders are matched with their corresponding numerical orders for all the modes.

The phases $\theta_{\rm g}$ and $\theta_{\rm p}$ 
%provide a way to classify the mixed modes from observations as well. They 
depict the phase variation that results from mode coupling while $\phi_{\rm p}$ represents the phase deviation, other than $\theta_{\rm p}$, brought in by the asymptotic analysis. Similar to the $\theta_{\rm g}$ and $\theta_{\rm p}$, $n_{\rm g}$ and $n_{\rm p}$ are related to $K$, thus to the mode frequencies which can be obtained from observations. Although the analysis introduced so far is done with modes computed using the Cowling approximation, we can apply the asymptotic phase-shift method to observations because we would be dealing generally with high-order observed modes and also the acoustic behaviour in the outer parts of the star is less affected by the perturbation to the gravitational potential. Therefore, in next section we apply the asymptotic phase-shift method to an observed \kepler\ star.

\section{Application to observed frequencies} \label{sc:obs}
The integral of $K$ provides us a useful tool to explore the interior structure of the star. It can be gained easily from models. In this section we discuss how to relate it to the observable parameters. \cite{jcd12} discussed the approximated versions of $K$ in the p and g cavities through a simple asymptotic analysis. In the p cavity for high order and low degree modes with $\omega^2 \gg N^2$ and $\omega^2 \gg S_l^2$, the integral of $K$ can be approximated by
\begin{equation}
\int_{r_{\rm c}}^{r_{\rm d}} K {\rm d} r \simeq \omega \int_{r_{\rm c}}^{r_{\rm d}} \frac{1}{c} \left(1 - S_l^2 \right)^{1/2 }{\rm d} r 
\simeq \omega \int_{r_{\rm c}}^{r_{\rm d}} \frac{1}{c} \; {\rm d} r \;.
\label{eq:kint}
\end{equation}
Here the second approximately equal sign is valid except near the turning point where $\omega = S_l$.
The outer turning point $r_{\rm d}$ is, in most cases, very close to the photosphere. And when the inner turning point $r_{\rm c}$ is very close to the stellar centre, the last term of equation~\eqref{eq:kint} is related to the $\Dnu_{\rm as}$ defined by equation~\eqref{eq:dnu}. However, for a giant star $r_{\rm c}$ is not close to the centre. Therefore, for each $l$ we define a $\Delta \nu_{\mbox{\tiny int},l}$ by
\begin{equation}
  \Dnu_{\mbox {\tiny int},l} = \left(2 \int_{r_{\rm c}}^{r_{\rm d}} \frac{1}{c} \; {\rm d} r \right)^{-1}\;.
\end{equation}
Then equation~\eqref{eq:kint} becomes
\begin{equation}
\int_{r_{\rm c}}^{r_{\rm d}} K {\rm d} r \approx \frac{\pi \nu}{\Dnu_{\mbox {\tiny int},l}} \;,
\label{eq:dnint}
\end{equation}
which can be substituted into equation~\eqref{eq:corthetap} to obtain
\begin{equation}
\nu \approx \Dnu_{\mbox {\tiny int},l} \left(n_{\rm p} + \frac{1}{\pi} \phi_{\rm p} + \frac{1}{\pi} \theta_{\rm p} \right) \;.
\label{eq:nu_int}
\end{equation}
For a radial mode $\theta_{\rm p}$ and the small separation $d_{nl}$ in equation~\eqref{eq:asyp} are zero; we then find that $\phi_{\rm p}/\pi$ in equation~\eqref{eq:nu_int} resembles $\epsilon_{\rm p}$. Since for a giant star, $r_{\rm c}$ is away from the stellar centre, $\Dnu_{\tiny {\rm int}, 0}$ is larger than $\Dnu_{\rm as}$, and, thus, we do not expect the two phases to be exactly the same, but still, we can assume that $\phi_{\rm p}$ is independent of $l$, just as $\epsilon_{\rm p}$ is. With this in mind, we can use the radial modes to determine $\phi_{\rm p}$ through
\begin{equation}
\phi_{\rm p} = \pi \left( \frac{\nu_{0}}{\Dnu_{\tiny {\rm int}, 0}} - n_{\rm p} \right) ,
\label{eq:phi_final}
\end{equation} 
where $\nu_{0}$ is the frequency  for $l = 0$ modes.
For a non-radial mode, $r_{\rm c}$ is even farther away from the centre which results in even larger $\Dnu_{\mbox {\tiny int},l}$, but $\phi_{\rm p}$ still does not depend on $l$ and this independence will be confirmed later. \cite{mos13} expressed the p mode frequency asymptotically, based on \cite{tas80}, including a second-order term $A_{\rm as}$ varying as $n^{-1}$:
\begin{equation}
\nu_{n,0} = \left( n + \epsilon_{\rm as} + \frac{A_{\rm as}}{n} \right) \Dnu_{\rm as},
\label{eq:asy_mosser}
\end{equation}
with a constant offset $\epsilon_{\rm as}$. Equation~\eqref{eq:asy_mosser} reproduces the curvature of the radial ridge very well in the observed frequency range. We fit equation~\eqref{eq:asy_mosser} to all of our models, and found for less-evolved models the curvature is not so pronounced and the frequencies of the high-order radial modes do not follow the form of $n^{-1}$. For RGB models the second-order term plays the same role as $\phi_{\rm p}$ and predicts the mode frequencies excellently, though the values of $A_{\rm as}$ and $\epsilon_{\rm as}$ differ from what \cite{mos13} found. Details about the fitting work are given in the Appendix.

Similarly, after comparing the expression of $K$ in the g cavity, we can also define $\Delta \Pi_{{\rm int},l}$ by
\begin{equation}
  \Delta \Pi_{{\rm int},l} = \left( \frac{\nu}{\pi} \int_{r_{\rm a}}^{r_{\rm b}} K {\rm d} r \right)^{-1}\;.
\label{eq:dpint}
\end{equation}
$\Dnu_{\mbox {\tiny int},l}$ and $\Delta \Pi_{\mbox {\tiny int},l}$ are theoretical frequency and period spacings defined by the integrals. They enable us to estimate the integrals of $K$ in each cavity for real stars, from observations, once we understand their relations with the observed spacings, $\Dnu_{\rm obs}$ and $\Delta \Pi_{\rm obs}$.  Since in observations dipolar mixed modes are predominantly detected, we will search for the relations between the spacings for dipolar modes. For convenience, hereafter we will use  $\Dnu_{\mbox {\tiny int}}$ and $\Delta \Pi_{\mbox {\tiny int}}$ to refer to $\Dnu_{\tiny {\rm int}, 1}$ and $\Delta \Pi_{\mbox {\tiny int}, 1}$, respectively.

In order to find these relations, we computed a large number of evolution sequences with masses ranging from 1.0 to 1.4 M$_{\sun}$, with chemical compositions [Fe/H] from $-0.5$ to $0.5$, and with mixing-length parameters from $1.5$ to $2.5$. The physics was kept the same as in models $M_{\rm s}$ and $M_{\rm r}$. More than ten thousand models were considered over different evolutionary stages, from the zero-age main sequence to late RGB, but the relations were searched mainly for SGB and RGB, with different spacing ranges. As an estimate of the observable $\Dnu_{\rm obs}$, we adopted the mean large separations of the mode frequencies, computed by performing a linear fit to several radial modes (normally 9 modes) around \numax ~\citep{jiang11}.  Moreover, the period \'{e}chelle diagram allowed us to estimate the observable $\Delta \Pi_{\rm obs}$ of the dipolar g modes or the g-m mixed modes \citep{bed11}. 
For the theoretical $\Dnu_{\mbox {\tiny int}}$ and $\Delta \Pi_{\mbox {\tiny int}}$, we integrated the $K$ defined in equation (\ref{eq:k}) for dipolar modes over the p- and g-mode propagation cavities, respectively.  This was also done for several dipolar modes around $\numax$ and then the theoretical period spacings are defined as the mean values of those obtained for each mode.

It should be noted that for some early SGB models the inadequate estimations of $K$ makes the boundaries of the two propagation regions ambiguous in the asymptotic analysis, which leads to deviations in the estimations of $\Dnu_{\mbox {\tiny int}}$ and $\Delta \Pi_{\mbox {\tiny int}}$. In that case, we calculated the integrals of $K$ for the quadrupolar modes in which the Lamb frequencies are larger than those for the dipolar modes, making the ambiguity of the propagation regions disappear, to obtain $\Dnu_{\mbox {\tiny int}}$ and $\Delta \Pi_{\mbox {\tiny int}}$. According to equation~\eqref{eq:dp} the period spacing is inversely proportional to $L$ and hence the spacings of the $l = 1$ and $2$ modes satisfy the relation as $\Delta \Pi_1 \approx \sqrt{3} \Delta \Pi_2$ which was verified by our models without ambiguous propagation regions, the difference between the two being within 1\,s.
However, the resulting differences between the frequency separations of $l$ = 1 and 2 modes can reach as high as 10\%; therefore for the computations of $\Dnu_{\mbox {\tiny int}}$ we simply ruled out these early SGB models that only take up a very small percentage of the whole set.
We then could compare $\Dnu_{\mbox {\tiny int}}$ and $\Delta \Pi_{\mbox {\tiny int}}$ with $\Dnu_{\rm obs}$ and $\Delta \Pi_{\rm obs}$, respectively, to link the integrals of $K$ with observations.        

We found the relation between $\Dnu_{\rm obs}$ and $\Dnu_{\mbox {\tiny int}}$ could be described by a simple linear fit as $\Dnu_{\mbox {\tiny int}}=1.089 \Dnu_{\rm obs} + 0.327$. The relation presented by \cite{mos13} does not fit our RGB models quite well and their slope of the linear fit is smaller than what we derived, because our fitting is performed for dipolar modes rather than radial modes and $\Dnu_{\mbox {\tiny int}}$ is larger than $\Dnu_{\mbox {\tiny int}, 0}$.
We also fitted $\Dnu_{\tiny {\rm int}, 0}$ to $\Dnu_{\rm obs}$ for the purpose of estimating $\phi_{\rm p}$ using equation~\eqref{eq:phi_final} and the relation is $\Dnu_{\mbox {\tiny int}, 0}=1.062 \Dnu_{\rm obs} + 0.083$ for which the slope is still greater than what was found by \cite{mos13}.

\begin{figure}
\resizebox{1.0\hsize}{!}{\includegraphics{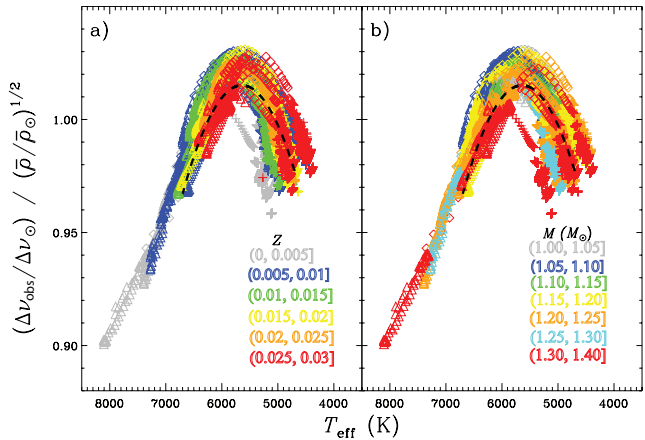}}
\caption{Ratio between $\Dnu_{\rm obs} / \Dnu_{\sun}$ and $(\bar \rho / \bar \rho_{\sun})^{1/2}$,
where $\bar \rho$ and $\bar \rho_\odot$ are the mean densities of the star and the Sun, respectively, as a function of effective temperature for all our models. $\Dnu_{\rm obs}$ are calculated from radial modes.
The ratio is presented in a) different metallicity and b) different masses with different colours. MS, SGB and RGB models are indicated by cross, diamond and triangle signs, respectively. The dashed black line shows the function given by equation (5) in \protect\cite{whi11}.}
\label{fg:scaling}
\end{figure}

With our computed $\Dnu_{\rm obs}$, we are also able to test the
scaling relation $\Dnu_{\rm obs} \propto (\bar \rho)^{1/2}$ where $\bar \rho$ is the mean density of the star.
Larger deviations from this relation are generally found for larger-mass and evolved stars than for lower-mass MS stars, and the deviations predominantly depend on the effective temperature \citep{whi11}.
The dependence of the scaling relation on effective temperature with our models is shown in Fig.~\ref{fg:scaling}. We do not find significant metallicity nor mass dependence. The effective temperature dependence found from our models basically matches what is given by equation (5) in \cite{whi11}, but the deviations of our models scatter within a range of 0.03.

Similarly, a linear fit was also performed to find the relation between the period spacings $\Delta \Pi_{\rm obs}$ and their theoretical counterparts $\Delta \Pi_{\mbox {\tiny int}}$. However, only models with a sufficient number of g modes were taken into consideration, which ruled out a few early SGB models. After performing a universal fit to all the models, we found the fit deviating substantially for the high-period-spacing models. This required us to do the fitting for SGB and RGB models separately. For stars with masses below 1.5 M$_{\sun}$, they pass the SGB and start climbing RGB when $(\Dnu_{\rm obs} / 36.5  \, \muHz) (\Delta \Pi_{\rm obs} / 126 \, \rm{s}) < 1$ \citep{mos14}. We followed this criterion and separated our models into two groups, so that $98\%$ of our models were considered as more evolved RGB models. The two model groups were fitted separately. The slope of the fit to the RGB is also very close to one and the relation is $\Delta \Pi_{\rm int, RG} = 1.004\Delta \Pi_{\rm obs} - 0.133$, while the relation for the SGB is $\Delta \Pi_{\rm int, SG} = 1.144\Delta \Pi_{\rm obs} - 14.69$. The latter relation is more suited for the models in the transition zone between subgiant and red-giant models. The relations between theoretical and observed values found so far allow us to proceed with the analysis of observational data. It should be noted that, to make this realistic, the results discussed so far in the current section are based on frequencies computed for the full fourth-order system.

\begin{figure}
\resizebox{1.0\hsize}{!}{\includegraphics{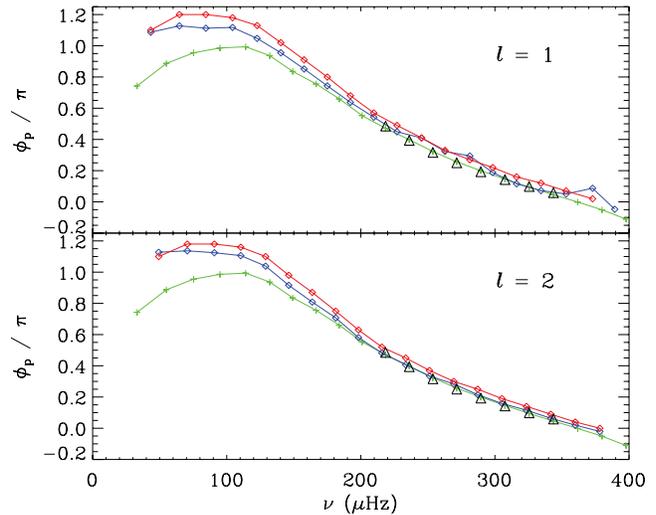}}
\caption{$\phi_{\rm p}$ obtained from three different methods for model $M_{\rm r}$ modes. $\phi_{\rm p}$ obtained by fitting the eigenfunction $\xi_{\rm r}$ as described in Section~\ref{sc:phi_p} is in red, using the integral of $K$ with equation~\eqref{eq:corthetap} in blue and using radial modes with equation~\eqref{eq:phi_final} in green. The black triangles correspond to the second-order term $A_{\rm as} / n + \epsilon_{\rm as}$ introduced in equation~(6) of \protect\cite{mos13}. The abscissa shows the frequency. }
\label{fg:phi_intVSphi}
\end{figure}

Since eigenfunctions cannot be obtained directly from observations, it is impossible to estimate $\phi_{\rm p}$ using the method mentioned in Section~\ref{sc:phi_p}. However, $\phi_{\rm p}$ can still be approximated using the radial modes and equation~\eqref{eq:phi_final}. We applied the two different methods, fitting $\xi_{\rm r}$ and using equation~\eqref{eq:phi_final}, to estimate $\phi_{\rm p}$ for radial as well as non-radial modes. The results for model $M_{\rm r}$ are shown in Fig.~\ref{fg:phi_intVSphi}. We used $\nu_0$ and $\Dnu_{\mbox {\tiny int}, 0}$ of the radial modes with equation~\eqref{eq:phi_final} to obtain the green curves which are identical in both panels. The blue curve is obtained by using the integral of $K$ with equation~\eqref{eq:corthetap} and the red ones present $\phi_{\rm p}$ obtained by fitting the eigenfunctions. The blue and red curves are computed for non-radial p-m mixed modes whose $\theta_{\rm p}$ is shown to be small by Table~\ref{tb:msg} and hence is negligible. The figures show that the red curves are very similar for different degree modes, which confirms that $\phi_{\rm p}$ is independent on $l$. 
More importantly, the fact that the green curve is a reasonable approximation to the other two curves demonstrates that the $\phi_{\rm p}$ obtained with equation~\eqref{eq:phi_final}, for radial modes, can be used also for non-radial modes, especially for those modes with frequencies around \numax\ ($\sim 253 \, \muHz$ for $M_{\rm r}$).
In Table~\ref{tb:msg}, we also give the values $\phi_{\rm p}^\prime$ of
$\phi_{\rm p}$ obtained by interpolating in the green curve to
the corresponding frequencies of p-m mixed modes.
Since the radial order for radial modes is generally uniquely defined
also for observed modes this technique provides an estimate of $\phi_{\rm p}$
also for the analysis of observations.
We use this in the following analysis.

\begin{figure}
\resizebox{1.0\hsize}{!}{\includegraphics{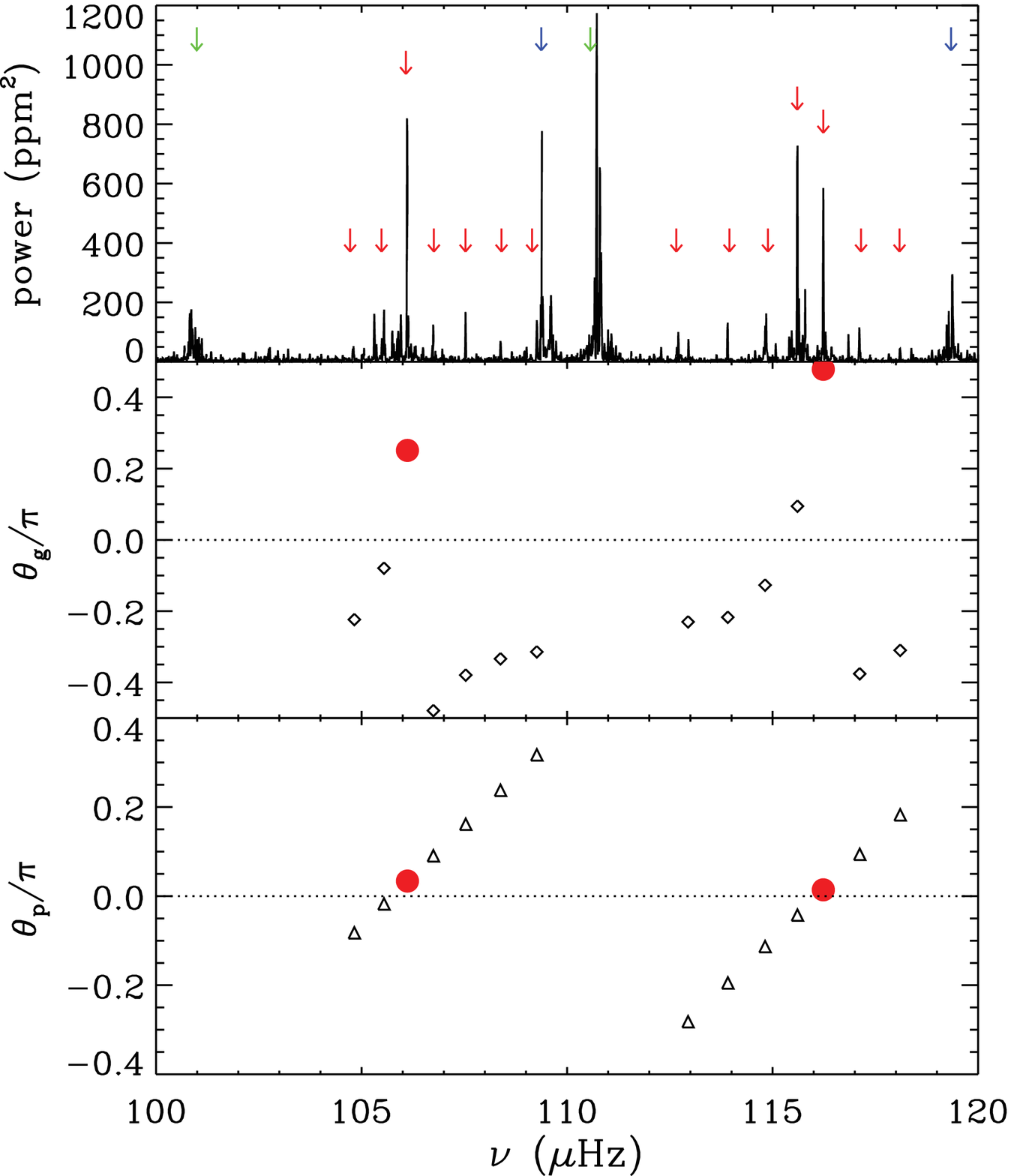}}
\caption{\textit{Top}: Power spectrum of the KIC 3744043, with superimposed mode identification provided by \protect\cite{mos12a}. Red arrows indicate the forests of dipolar mixed modes, while green and blue arrows show the positions of radial and quadrupole modes, respectively. The power spectrum only covers two $\Dnu_{\rm obs}$ range, with $\numax$ of $110.9 \, \muHz$, also obtained by \protect\cite{mos12a}, located in the centre of the spectrum. \textit{Centre}: $\theta_{\rm g}$ for the dipolar mixed modes, as a function of the frequency. Modes with $\theta_{\rm g}$ close to 0 preserve more g-mode character. On the other hand, modes with $\theta_{\rm g}$ away from 0 are more affected by p-mode character. \textit{Bottom}: The $\theta_{\rm p}$ including the phase offset $\phi_{\rm p}$. The most p-mode-like modes in each cluster are highlighted by red dots in the last two plots.}
\label{fg:psp}
\end{figure}

We applied the phase-shift approach to the ascending-branch red-giant star KIC 3744043 that was observed by the \kepler\ space mission. This star was analysed by \cite{mos12a} for mixed-mode study. They found that the $\Dnu_{\rm obs}$ and $\Delta \Pi_{\rm obs}$ for this star are 9.90\,$\muHz$ and 75.98\,s, respectively. Using the linear relations between the theoretical spacings and the observed ones for $l = 1$ modes, we approximated $\Dnu_{\mbox {\tiny int}}$ and $\Delta \Pi_{\mbox {\tiny int}}$ to be 11.11\,$\muHz$ and 76.15\,s. A part of the power spectrum of KIC 3744043 is plotted in the top panel of Fig.~\ref{fg:psp}, covering  a $2\Dnu_{\rm obs}$-wide frequency range, with $\numax$ of 110.9 $\muHz$ located at the centre. 
In addition to two radial modes and two $l=2$ p modes, two groups of $l=1$ mixed modes are clearly illustrated. 
We estimated the degrees of mixing for the dipolar modes by calculating $\theta_{\rm g}$ and $\theta_{\rm p}$ with $\phi_{\rm p}$, using equations~\eqref{eq:eg} and \eqref{eq:corthetap}, respectively. The integrals of $K$ in the p- and g-mode cavities were estimated from the observed individual frequencies, $\Dnu_{\mbox {\tiny int}}$ and $\Delta \Pi_{\mbox {\tiny int}}$ with equations~\eqref{eq:dnint} and \eqref{eq:dpint}.
$\Dnu_{\mbox {\tiny int}, 0}$ was estimated as 10.60\,$\muHz$ from the relation between $\Dnu_{\mbox {\tiny int}, 0}$ and $\Dnu_{\rm obs}$ mentioned above.
For each group of dipolar p-m modes $\phi_{\rm p}$, assumed the same for all modes
in the group, was then estimated from applying~\eqref{eq:phi_final} to the
radial mode with frequency just below the group,
and this
was then used to calculate $\theta_{\rm p}$ as described by equation~\eqref{eq:corthetap}.

The lower two panels in Fig.~\ref{fg:psp} show  
the $\theta_{\rm g}$ and the $\theta_{\rm p}$ resulting from
this analysis.
As discussed above, these p-m mixed modes possess the p-mode character through coupling. The degree of mixing is reflected in $\theta_{\rm g}$ and $\theta_{\rm p}$. In this case, the most p-m like mixed modes have the largest absolute values of $\theta_{\rm g}$ in their local mode clusters and they are located the farthest away from the dashed line indicating the position of 0 while their $\theta_{\rm p}$ are the closest to 0. Normally, the p modes have much larger amplitude than g modes, which makes them much easier to observe.
This is true in the power spectrum of KIC 3744043. In the first dipolar mode cluster around 106 $\muHz$, the third mode has the smallest absolute $\theta_{\rm p}$ value as well as obviously the highest peak, and therefore it is the most p-m like mode in this cluster.
Based on $\theta_{\rm g}$ the third mode in the cluster indicates also the
smallest departure from the g-mode behaviour.
The second cluster of dipolar modes also proves the validity of using the phase shift. The amplitudes of the most p-m like modes are usually the highest among the mode cluster%
\footnote{We note that in the second cluster the most p-m like mode is assigned to the mode with the second highest peak based on the vales of $\theta_{\rm g}$, but the $\theta_{\rm p}$ of this mode is only a little closer to 0 than that of the mode with the highest peak.},
but it is not possible to distinguish the rest of the mixed modes, which have similar amplitudes in the cluster, in terms of their mixed characters from the power spectrum. One might do so by plotting the period spacings between the modes with consecutive radial orders. However, consecutive modes are not always detectable from the power spectrum. In this case, the phase-shift method gives a better way to classify mixed modes directly and easily. 

We can also identify the radial order for these dipolar modes. Since $\phi_{\rm p}$ is already known, we just need to estimate the integrals of $K$. These are estimated from the observed frequencies and $\Dnu_{\mbox {\tiny int}}$ or $\Delta \Pi_{\mbox {\tiny int}}$, respectively, which, in turn, are computed  from $\Dnu_{\rm obs}$ and $\Delta \Pi_{\rm obs}$ using the linear relation discussed above. With $\phi_{\rm p}$ and the integrals in hand, we then compute $n_{\rm p}$ and $n_{\rm g}$  by applying the method described in Section~\ref{sc:order}. We have obtained the mode orders for the two groups of mixed modes in KIC 3744043 to be 9 and 10, which are identical to what \cite{mos12a} found. The g-mode radial orders $n_{\rm g}$ are from 125 to 110.

\section{Conclusion}

In SGB and RGB stars the degrees of mixing for some modes differs drastically from that of their neighbouring modes. For these modes their frequencies also vary significantly from the regular asymptotic p-mode pattern in SGB and asymptotic g-mode pattern in RGB stars. These modes theoretically satisfy the conditions set by equations~\eqref{eq:gcon} and \eqref{eq:pcon_cor} for mixed modes. In this paper, we discussed a method of utilizing the phase shifts $\theta_{\rm p}$ and $\theta_{\rm g}$ as a tool to distinguish the degree of mixing for adjacent mixed modes.

We computed $\theta_{\rm p}$ and $\theta_{\rm g}$ values for the mixed modes in a SGB and a RGB model. For both cases, $\theta_{\rm p}$ or $\theta_{\rm g}$ are around 0 when the mode suitably satisfies the eigenfunction conditions for pure p or g modes, respectively. When the modes meet the condition for mixed modes, the absolute values of $\theta_{\rm p}$ and $\theta_{\rm g}$ shift away from zero to different extents, meaning the mode gains p-mode character if it was a g mode before and vice versa. Therefore, the absolute value of $\theta_{\rm p}$ and $\theta_{\rm g}$ can be used to classify mixed modes and to determine their degrees of mixing. We applied this method to the red-giant star KIC 3744043 observed by \kepler. From the power spectrum of this star, it is not possible directly to classify the p-mode character of the dipolar mixed modes in the same frequency cluster. With the help of $\theta_{\rm p}$ and $\theta_{\rm g}$, we could accomplish that by measuring how far away their $\theta_{\rm g}$ values are from zero or how close their $\theta_{\rm p}$ values are to zero, once we obtained the $\Dnu_{\rm obs}$ and $\Delta \Pi_{\rm obs}$ from the power spectrum. This method works well for high-order observable modes.

Another purpose of computing $\theta_{\rm p}$ and $\theta_{\rm g}$ is to identify the radial orders of p- and g-mode components that mix into a mixed mode. The mixed-mode radial order indicates the number of radial nodes in the eigenfunction in the radial direction. In order to have an accurate and continuous numbering of mixed modes, which is given in theoretically computed modes, we use the definition $n = n_{\rm p} - n_{\rm g}$, where $n_{\rm p}$ and $n_{\rm g}$ are the pressure and gravity radial orders and can be obtained from equations~\eqref{eq:corthetap} and \eqref{eq:eg}.
We do note, however, that the determination of radial orders is tricky for those modes undergoing strong mixing. For those modes the absolute values of  $\theta_{\rm g}$ and $\theta_{\rm p}$ are around 0.5, making their radial orders hard to decide upon.
A more basic difficulty results from the approximations used in the asymptotic analysis, which cause clear deviations from the numerical results for low-order modes.
After comparing the numerical and asymptotical eigenfunctions, we found that the deviations of $n_{\rm p}$ can be corrected by a phase offset $\phi_{\rm p}$ estimated by matching the asymptotic eigenfunction to the numerical one.
For observational data it is obviously not possible to estimate $\phi_{\rm p}$ from the eigenfunctions;
here we linked $\phi_{\rm p}$ to the radial mode frequency and $\Dnu_{\mbox {\tiny int}, 0}$, which provides an alternative way to estimate $\phi_{\rm p}$.
The two methods used in our analysis to estimate $\phi_{\rm p}$ are consistent with each other. The first way with the eigenfunctions is optimal for models when the eigenfunctions are available.
The second procedure, using the radial modes, gives good estimates of $\phi_{\rm p}$ for modes around \numax\ and hence is suitable for observational data.
Using this technique we obtained reasonable estimates for the phase shifts and the mode orders for the red-giant star KIC 3744043.

Knowing the pressure and gravity radial orders gives us a better understanding of the contributions to mixed modes from the p- and g-mode character. The phase shift provides us an alternative way to identify the mode other than the normal frequency and period \'{e}chelle diagrams, especially to identify the mode mixture for a mixed mode, with an important byproduct of the estimations of mode orders using only mode frequencies and the spacings. We should note that the models used in this work both have relatively thick intermediate evanescent regions. It is essential that this region is not assumed to be always thin. A thin evanescent region means strong coupling between acoustic- and gravity-wave and can push the coupling strength approaching 1.  Since the eigenvalue condition of equation~\eqref{eq:fullcondi} still holds the validity when the evanescent region is thin \citep{tak16b}, we assume the phase shifts method also works in that situation, which will be thoroughly examined in future work.

\section{Acknowledgements}

The authors wish to thank the entire \kepler\ team, without whom these results would not be possible. Funding for this Discovery mission is provided by NASA's Science Mission Directorate. Special thanks to the referee for pertinent comments on earlier versions of the manuscript which greatly improved the presentation. The research is supported by he European Community Programme ([FP7/2007-2013]) under grant agreement No. FP7-SPACE-2012-312844 (SPACEINN) and by Funda\c c\~ao para a Ci\^encia e a Tecnologia (FCT) through national funds (UID/FIS/04434/2013) and by FEDER through COMPETE2020 (POCI-01-0145-FEDER-007672). MC also acknowledges support from FCT through Investigador FCT contractIF/00894/2012/CP0150/CT0004. This work was also supported in part by the ASTERISK project (ASTERoseismic Investigations with SONG and \kepler) funded by the European Research Council (grant agreement no.: 267864). Funding for the Stellar Astrophysics Centre is provided by the Danish National Research Foundation (grant agreement no.: DNRF106).

\clearpage

\label{lastpage}

\clearpage

\appendix
\section{The asymptotic second-order term}
\label{app:appA}
\begin{figure}
\resizebox{1.0\hsize}{!}{\includegraphics{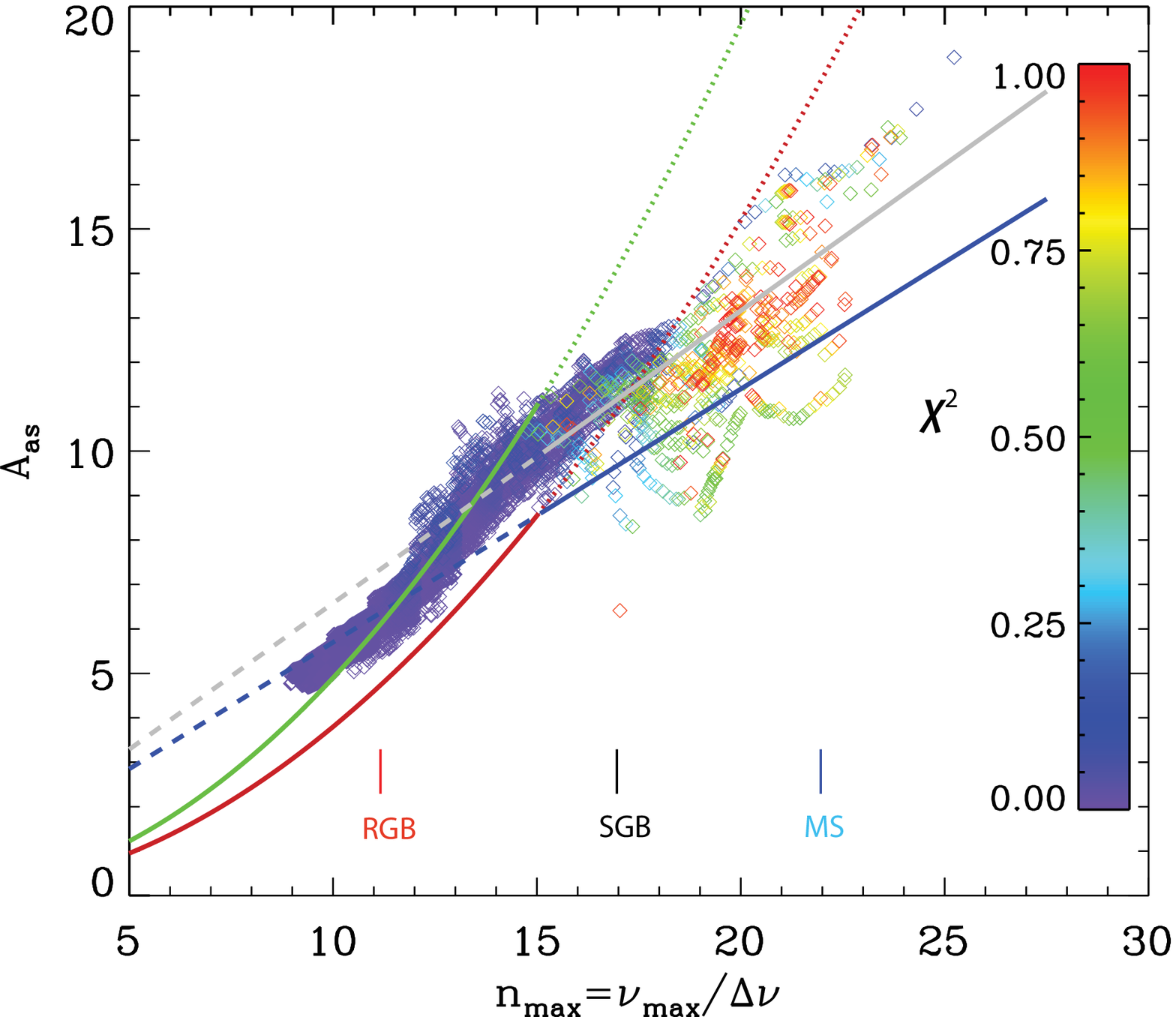}}
\caption{The second-order asymptotic term $A_{\rm as}$ obtained by fitting equation~(6) of \protect\cite{mos13} to all our model frequencies, as a function of $n_{\rm max} = \nu_{\rm max} / \Dnu_{\rm obs}$. The colour code of the symbols provides the $\chi^2$ of the fit. The red and blue lines correspond to the fit in $n_{\rm max}^{-1}$ and $n_{\rm max}^{-2}$, respectively, adopted directly from \protect\cite{mos13}, while the green and grey lines correspond to the fit in our analyse. The colours of the vertical lines indicate the different regime of evolution stages.}
\label{fg:as_mosser}
\end{figure}

\begin{figure}
\resizebox{1.0\hsize}{!}{\includegraphics{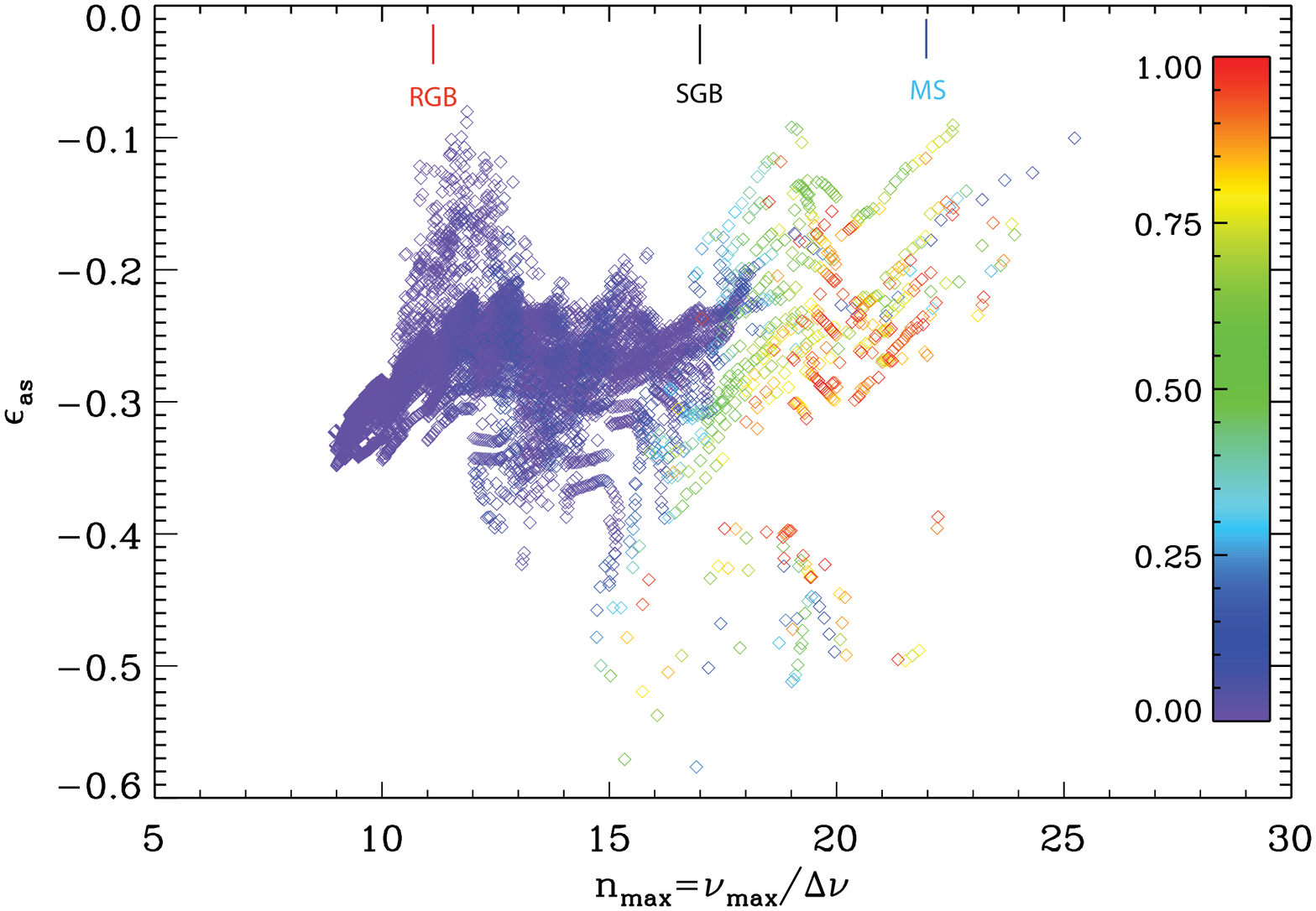}}
\caption{Same as Fig.~\ref{fg:as_mosser}, for the asymptotic offsets $\epsilon_{\rm as}$.}
\label{fg:epsilon_mosser}
\end{figure}

\cite{mos13} proposed that the measurement of the curvature of the radial ridges in the \'{e}chelle diagram provides a correction to the observed value of large separation. They derived a link between the asymptotic large separation $\Dnu_{\rm as}$ and its observed counterpart using series of radial-mode frequencies in published analyses of stars observed by \kepler\ or \corot. The curvature of the observed radial ridge reproduces a second-order term in the asymptotic expression for p modes, which was taken into consideration in the derivation of the asymptotic large separation. In our analysis, the effect of the second-order term is included in the frequency-dependent $\epsilon_{\rm p}$ or the phase $\phi_{\rm p}$. We make use of the abundant sample of models,
computed to relate $\Dnu_{\mbox {\tiny int}}$ and $\Delta \Pi_{\mbox {\tiny int}}$ to $\Dnu_{\rm obs}$ and $\Delta \Pi_{\rm obs}$ (cf. Section \ref{sc:obs}),
covering large ranges in mass and chemical composition to test the second-order term $A_{\rm as} / n$ in the asymptotic regime by fitting equation~\eqref{eq:asy_mosser} to our model frequencies. \cite{mos13} used radial modes in the frequency range where modes have appreciable amplitudes. In order to resemble their work, we also adopt the modes around $n_{\rm max}$, typically using 7 modes in total for a given model. The significance of the fit is demonstrated by the value of $\chi^2$ defined as:
\begin{equation}
\chi^2 = \sum \frac{(\nu - \nu^\prime)^2}  {\sigma}.
\end{equation}
where $\nu^{\prime}$ are the frequencies from the fitting and $\sigma$ is set to $1 \,\muHz$ as we do not have errors for model frequencies. The results are shown in Figs.~\ref{fg:as_mosser} and \ref{fg:epsilon_mosser} where $A_{\rm as}$ and $\epsilon_{\rm as}$ are plotted as functions of $n_{\rm max}$. Only results with $\chi^2 < 1$ are displayed, which represent acceptable fits. We note that the less-evolved models have less pronounced curvature ridges to which the fitting is very sensitive. This leads to a tendency for $\chi^2$ to increase as $n_{\rm max}$ becomes larger. On the other hand, the fits for the RGB models show that a $n^{-1}$ dependence in equation~(6) of \cite{mos13} is workable for red giant stars, at least in the observed frequency range. Our resulting $A_{\rm as}$ are generally larger than what \cite{mos13} found, while $\epsilon_{\rm as}$ is smaller than their results which are around $1/4$. 
This may be related to the fact that \citet{mos13} analysed observed frequencies, differing from our model frequencies owing to the neglect of near-surface effects in the modelling.
%Since it is important that \cite{mos13} analyse observed modes and thus the surface errors are relevant. The missing of the surface errors in our models might account for the difference in $A_{\rm as}$ and $\epsilon_{\rm as}$.
Our fit of $A_{\rm as}$ in $n_{\rm max}$ gives the relation for RGB models 
\begin{equation}
A_{\rm as, RG} = 0.049 n_{\rm max}^2,
\end{equation}
and for MS models
\begin{equation}
A_{\rm as, MS} = 0.66 n_{\rm max}.
\end{equation}

We also applied the fit of equation (\ref{eq:asy_mosser}) to the radial-mode
frequencies of model $M_{\rm r}$,
%The second-order term $A_{\rm as} / n$ plus the constant term $\epsilon_{\rm as}$ introduced in equation~(6) of \cite{mos13} are also plottedi
plotting the result for several radial modes around $\nu_{\rm max}$ in Fig.~\ref{fg:phi_intVSphi}.
%The two terms are obtained in the way mentioned before, by fitting equation~(6) of \cite{mos13} to our models.
The fit reproduces the mode frequencies very well, with a $\chi^2$ of 0.19. Therefore, it is not surprising to see that they overlap the $\phi_{\rm p}$ computed using equation~\eqref{eq:phi_final}.
A more detailed comparison between our work and \cite{mos13}
%requires another thorough project and
is beyond the scope of this paper, however.

\end{document}